\newif\iffigs\figstrue

\documentclass[12pt]{article}
\usepackage{amstex,epsf,amssymb}
\DeclareFontFamily{U}{rsf}{}
\DeclareFontShape{U}{rsf}{m}{n}{
  <5> <6> rsfs5 <7> <8> <9> rsfs7 <10-> rsfs10}{}
\DeclareMathAlphabet\Scr{U}{rsf}{m}{n}

\def\pplogo{\vbox{\kern-\headheight\kern -15pt
\halign{##&##\hfil\cr&{
\ppnumber}\cr\rule{0pt}{2.5ex}&\ppdate\cr}
}}
\makeatletter
\def\ps@firstpage{\ps@empty \def\@oddhead{\hss\pplogo}%
  \let\@evenhead\@oddhead 
}
\def\maketitle{\par
 \begingroup
 \def\thefootnote{\fnsymbol{footnote}}
 \def\@makefnmark{\hbox
 to 0pt{$^{\@thefnmark}$\hss}}
 \if@twocolumn
 \twocolumn[\@maketitle]
 \else \newpage
 \global\@topnum\z@ \@maketitle \fi\thispagestyle{firstpage}\@thanks
 \endgroup
 \setcounter{footnote}{0}
 \let\maketitle\relax
 \let\@maketitle\relax
 \gdef\@thanks{}\gdef\@author{}\gdef\@title{}\let\thanks\relax}
\makeatother

\def\P{{\mathbb P}}

\def\R{{\mathbb R}}
\def\Z{{\mathbb Z}}

\def\Area{\operatorname{Area}}
\def\Vol{\operatorname{Vol}}

\def\Sl{\operatorname{SL}}
\def\GO{\operatorname{O{}}}

\def\Spin{\operatorname{Spin}}

\def\CY{Calabi--Yau}
\def\MW{Mordell--Weil}

\def\cM{{\Scr M}}

\def\cD{{\Scr D}}

\def\cX{{\Scr X}}

\def\cMc{{\hfuzz=100cm\hbox to 0pt{$\;\overline{\phantom{X}}$}\cM}}
\def\barcD{{\hfuzz=100cm\hbox to 0pt{$\;\overline{\phantom{X}}$}\cD}}
\def\ff#1#2{{\textstyle\frac{#1}{#2}}}

\def\spnh{\Spin(32)/\Z_2}

\def\HS#1{{\mathbb{F}}_{#1}}

\begin{document}
\setcounter{page}0
\title{\LARGE M-Theory Versus F-Theory Pictures\\ of the Heterotic
String\\[10mm]} 
\author{
Paul S. Aspinwall\\[0.7cm]
\normalsize Dept.~of Physics and Astronomy,\\
\normalsize Rutgers University,\\
\normalsize Piscataway, NJ 08855\\[10mm]
}
\def\ppnumber{\vbox{\baselineskip14pt\hbox{RU-97-55}
\hbox{hep-th/9707014}}}
\def\ppdate{June 1997} \date{}

{\hfuzz=10cm\maketitle}

\def\Large{\large}
\def\LARGE{\large\bf}

\vskip 1cm

\begin{abstract}
If one begins with the assertion that the type IIA string compactified
on a K3 surface is equivalent to the heterotic string on a four-torus
one may try to find a statement about duality in ten dimensions by
decompactifying the four-torus. Such a decompactification renders the
K3 surface highly singular. The resultant K3 surface may be analyzed
in two quite different ways --- one of which is natural from the point
of view of differential geometry and the other from the point of view
of algebraic geometry. We see how the former leads to a ``squashed K3
surface'' and reproduces the Ho\v{r}ava--Witten picture of the
heterotic string in M-theory. The latter produces a ``stable
degeneration'' and is tied more closely to F-theory. We use the
relationship between these degenerations to obtain the M-theory
picture of a point-like $E_8$-instanton directly from the F-theory
picture of the same object.
\end{abstract}

\vfil\break


\section{Introduction}

As is now well-known, the original multitude of dualities involving
various string theories and M-theory are highly interrelated and can
be reduced down to some minimal set from which all others may be
deduced. For example, Sen \cite{Sen:uni} showed that the known
dualities could be reduced to T-dualities (i.e., dualities between
weakly coupled strings that can be deduced from world-sheet methods)
and the duality between the type I open string and the $\spnh$
heterotic string in ten dimensions. In this approach M-theory is
defined as the strongly-coupled limit of the type IIA
string. We will also adopt this definition of M-theory here even though
there has been recent progress in a more intrinsic approach to
M-theory \cite{BFSS:M}.

It is, of course, largely a question of taste as to which string
duality is to be considered as more fundamental than another. In this
paper we will take the duality between the type IIA string on a K3
surface and the heterotic string on a four-torus to be
fundamental. One motivation for this is that compactification of
closed string theories on complicated objects, such as \CY\ manifolds,
is currently better-understood than that for open strings.

If one is to take a six-dimensional pair of string compactifications
as a fundamental axiom of string duality then one needs to be able to
``decompactify'' this dual pair in order to make statements about
higher-dimensional theories. It is the geometry of this
decompactification process that we wish to discuss in this paper.

The majority of this paper will not be particularly original as most
of the results we discuss have appeared in various guises elsewhere,
especially in \cite{MV:F2,FMW:F}. Given the large number of dualities
at our disposal, any given problem may be solved in a large number of
ways. For example, the recent paper \cite{CJ:MF} covers similar
questions to those posed here but from a different viewpoint.
The main purpose of this paper will be to emphasize a particularly direct
relation between the geometry of F-theory and of M-theory especially
where the structure of the $E_8\times E_8$ heterotic string is
apparent. Each picture will turn out to be closely related to a
decompactification question.

The moduli space of a type IIA string on a K3 surface and the moduli
space of a heterotic string on a four-torus are understood and are
exactly the same. This allows us to know exactly which K3 surfaces
correspond to exactly which four-torus. See \cite{me:lK3} for a review
of these facts. We may then ask the question of what exactly happens
to the K3 surface as the 1-cycles in the four-torus are taken to be
very large. This should tell us how the type IIA string can be dual to
the heterotic string in more than 6 dimensions.

An important aspect about this process with respect to the moduli
space concerned is that this degeneration consists of going an
infinite distance in the moduli space. This is what makes the analysis
somewhat subtle.

We will illustrate the two types of limits we wish to consider in the
moduli space by examining the well-known moduli space of a two-torus
in section \ref{s:T2}. In section \ref{s:M} we will then squash a K3
surface as an application of one type of limit. This will reproduce
the Ho\v{r}ava--Witten picture \cite{HW:E8M} of the $E_8\times E_8$
heterotic string on a line segment. We will then use the other kind of
limit to produce the stable degeneration picture on F-theory in
section \ref{s:F}. Finally in section \ref{s:MF} we use the
relationship between the two degenerations to turn an F-theory picture
into an M-theory picture. The example studied will be that of
point-like $E_8$-instantons corresponding to 5-branes peeling off the
end of the M-theory line segment as in \cite{GH:E8i,SW:6d}.


\section{The Two-torus}  \label{s:T2}

Before we analyze the contortions our K3 surface will undergo in order
to decompactify the heterotic four-torus, we will begin with the
simplest example of an infinite-distance degeneration. This is that of
the two-torus, or elliptic curve.

The usual description of $T^2$ is that of the complex $z$-plane
divided out by the identifications
\begin{equation}
  z\mapsto z+1,\quad z\mapsto z+\tau,
\end{equation}
where $\tau$ is some complex number, $\mathrm{Im}(\tau)>0$. We do not
care about the overall 
size of the torus --- just the complex structure. In this case $\tau$
is only defined up to the $\Sl(2,\Z)$ transformations
\begin{equation}
  \tau\mapsto\frac{az+b}{cz+d}.
\end{equation}
This allows us to restrict $\tau$
to the well-known fundamental region
\begin{equation}
  -\ff12\leq\mathrm{Re}(\tau)\leq\ff12,\quad|\tau|\geq 1.
\end{equation}
So long as $\tau$ is finite within this region we clearly have a nice
smooth torus. In the limit where $\tau$ becomes
\begin{equation}
  \tau_\infty = \lim_{\beta\to\infty} \alpha + i\beta,
		\label{eq:tinf}
\end{equation}
where $\alpha$ and $\beta$ are real, the torus degenerates. This is
the object of interest.

Metrically it is clear what has happened to the torus if we inherit
the flat metric of the $z$-plane. One of the
1-cycles has become infinitely larger than the other. Since we didn't
fix the overall area of the torus, there are two obvious
possibilities. One is that one cycle remains fixed and the other
becomes infinite, or that one cycle remains fixed and the other
shrinks to zero size. 

One answer to the question of what the degenerate torus looks like
therefore is that it has become very ``skinny'' and shrunk itself down
to a circle. That is, the dimension has decreased from two real
dimensions to one real dimension. Note that the value of $\alpha$
in (\ref{eq:tinf}) is irrelevant in this limit.

This we take to be the differential geometers picture of the
degenerate torus. If one uses the flat metric on the torus to
determine the exact shape of the limit then the above picture is
unavoidable. As is well-known however, this is not the only way of
analyzing the problem. If one asks an algebraic geometer to write down
a torus, one might end up with the complex curve
\begin{equation}
  x_0^3 + x_1^3 + x_2^3 -3\psi\,x_0x_1x_2=0,
		\label{eq:cub}
\end{equation}
in $\P^2$. Now the complex structure is governed by $\psi$ up to the
equivalence 
\begin{equation}
  \psi\mapsto \omega\psi,\quad
  \psi\mapsto-\frac{2+\psi}{1-\psi},
\end{equation}
where $\omega=\exp(2\pi i/3)$.

Now, if $\psi\neq 1$, or equivalent, then this elliptic curve is
smooth and is just as good as description of a 2-torus as that given
earlier and one may deduce a finite value of $\tau$ corresponding to a given
value of $\psi$. It might therefore appear that $\tau=\tau_\infty$
corresponds to $\psi=1$. This is not literally true however. When
$\psi=1$, (\ref{eq:cub}) factorizes into
\begin{equation}
  (x_0+x_1+x_1)(x_0+\omega x_1+\omega^2x_1)(x_0+\omega^2x_1+\omega x_1)
      =0.  \label{eq:cub0}
\end{equation}
  Now each of the three factors in (\ref{eq:cub0}) is a linear
hyperplane in $\P^2$ and is therefore a sphere. Each pair of spheres
touch each other once. Thus we have a ring of three 2-spheres. 

This picture of the degenerated torus thus corresponds to pinching
down three parallel 1-cycles. This is quite different to the
differential geometers answer above. The dimension of the torus hasn't
changed at all in the algebraic geometers picture --- only a finite
number of 1-cycles have been shrunk down. In the differential
geometer's picture {\em all\/} the 1-cycles in a given class were
shrunk down to zero size.

The degeneration of the torus given by (\ref{eq:cub0}) is known as a
``semi-stable'' degeneration. The general idea of a semi-stable
degeneration is to write down a fibration
\begin{equation}
  \pi:\cX \to\Delta,
\end{equation}
where $\cX$ is a smooth algebraic variety of dimension one greater
than that required and $\Delta$ is some open subset of the complex
$t$-plane. The variety of interest will then be the fibre
$\pi^{-1}(t)$ which will be smooth everywhere except at $t=0$ which
will correspond to the degenerated space.

That is, a semi-stable degeneration of an elliptic curve corresponds
to the possible singular fibres of an elliptic fibration of a complex
surface.\footnote{One can also demand that the central fibre,
$\pi^{-1}(0)$, be a reduced divisor by allowing for suitable base
changes. See \cite{FM:BGOD} for more details.}
These were classified by Kodaira and have appeared frequently
in the context of F-theory, see, for example, \cite{MV:F2,me:lK3}.

The semi-stable degeneration is not particularly satisfactory for our
purposes as the answer is not unique. The idea of a ``stable''
degeneration is to find the ``most generic'' semi-stable degeneration
and deem it more fundamental than the others. In the case of elliptic
curves, the natural degeneration to choose is the one in which
precisely {\em one\/} cycle has been shrunk down. See, for example,
chapter one of \cite{FM:BGOD} for an explanation of this together with
the relevant references.

We have seen in this section that there are two natural models for a
degeneration of a torus. To the metrically minded, the torus shrinks
itself onto a circle. To the algebraically minded, the degeneration is
less severe and consists of just one cycle in the torus shrinking to a
point. 


\section{M-Theory}    \label{s:M}

The purpose of this section is to fulfill the promise made in
\cite{AM:Ud} that the ten-dimensional heterotic string can be
understood as M-theory on a squashed K3 surface. This will be consistent
with the Ho\v{r}ava--Witten picture \cite{HW:E8M} in that this
squashed K3 surface will be a finite line segment, $\mathcal{I}$,
(i.e., $S^1/\Z_2$). The general principles of this calculation were
outlined in \cite{MV:F2}. 
The Ho\v{r}ava--Witten picture or squashed K3 surface has also
appeared in discussions of M(atrix) theory --- see, for example,
\cite{BR:K3M,Gov:K3M}.

Our method will be that of the differential geometer who uses a metric
on the K3 surface to analyze the exact way in which it degenerates. We
will only consider the $E_8\times E_8$ heterotic string here. There
appears to be no obvious obstacle in applying this technique to the
$\spnh$ heterotic string although it will be technically harder. If an
answer for the $\spnh$ heterotic string could be found it would be
very interesting given the equivalence of the strongly-coupled $\spnh$
heterotic string to the type I string in ten dimensions.

Let us consider the heterotic string on the four-torus. We will
decompactify as follows. Firstly switch off all the Wilson lines to
give an $E_8\times E_8$ gauge symmetry. Then set the $B$-field to zero
and make the $T^4$ be given by a product of four orthogonal circles
of radius $\sqrt{\alpha}_i$, for $i=1,\ldots,4$. We may treat all four
directions on an equal footing by letting all four radii be
equal. We should have in mind then that we are going to ultimately set
$\alpha_1=\alpha_2=\alpha_3=\alpha_4$. Having said that, in order to
present the 
construction it will be easiest if we allow the notation to
distinguish between the various $\alpha_i$'s. We therefore treat the
$\alpha_i$'s as independent for most of the time.
To decompactify we let $\alpha_i\to\infty$. What
does all this do to the K3 surface of the dual type IIA string?

We may answer this question given the explicit form of the map between
the heterotic string's four-torus and the type IIA string's K3 surface
\cite{AM:K3p}. Most of the manipulations we will perform are explained
at length in \cite{me:lK3} (in particular, section 5.6) and so we will
be brief here. 
Recall that the moduli space of both the K3
surface and the four-torus in question is given by
\begin{equation}
  \cM \cong
\GO(\Gamma_{4,20})\backslash\GO(4,20)/(\GO(4)\times\GO(20)),
\end{equation}
which may be interpreted as the Grassmanian of space-like 4-planes, $\Pi$,
within $\R^{4,20}$, divided by the isometries of the even unimodular
lattice $\Gamma_{4,20}\subset\R^{4,20}$.

To achieve the $E_8\times E_8$ gauge symmetry we decompose
\begin{equation}
  \Gamma_{4,20}\cong \Gamma_8\oplus\Gamma_8\oplus\Gamma_{4,4},
	\label{eq:dec1}
\end{equation}
and impose that $\Pi$ is orthogonal to $\Gamma_8\oplus\Gamma_8$. 
This produces two $E_8$ quotient singularities in the K3 surface.
Let $w^1,\ldots,w^4$ be four null vectors in $\R^{4,4}$ and let
$w^*_1,\ldots,w^*_4$ be their duals. Together they generate
$\Gamma_{4,4}$. Let $\Pi$ be given by the span of the vectors
\begin{equation}
  w^*_i + \sum_j\psi_{ij}w^j,
\end{equation}
for some matrix $\psi$. One may then show (cf.\ section 5.6 of
\cite{me:lK3}) that the desired location in moduli space is given by
\begin{equation}
  \psi=\begin{pmatrix}\alpha_1&0&0&0\\0&\alpha_2&0&0\\0&0&\alpha_3&0\\
     0&0&0&\alpha_4\end{pmatrix},  \label{eq:psi}
\end{equation}
for $\alpha_i\to\infty$.

For the degeneration we wish to perform, we are free to decompose further
\begin{equation}
  \Gamma_{4,4}\cong \Gamma_{2,2}\oplus\Gamma_{2,2},
\end{equation}
and assert that $\Pi$ is spanned by two space-like 2-planes, $\Omega$
and $\mho$, each of which is a subspace of
$\Gamma_{2,2}\otimes_\Z\R$. In terms of the K3 surface this amounts to
asserting that we have an algebraic K3 surface whose complex structure
is determined by $\Omega$ and whose K\"ahler form and $B$-field is
given by $\mho$. The 24 elements of the cohomology of the K3 surface
can be divided into three sets:
\begin{enumerate}
  \item The 16 classes corresponding to the cycles shrunk to zero size
to produce the two $E_8$ singularities and hence the $E_8\times E_8$
gauge symmetry.
  \item The four classes spanning $\Gamma_{2,2}$ giving the ``quantum
Picard lattice'' which is the classical Picard lattice plus
$\Gamma_{1,1}$ generated by $H^0$ and $H^4$. Thus the Picard lattice is
$\Gamma_{1,1}$.
  \item The remaining transcendental lattice of 2-cycles given by
$\Gamma_{2,2}$.
\end{enumerate}
The fact that the Picard lattice is $\Gamma_{1,1}$ is sufficient to
show that the K3 surface is an elliptic fibration with a unique
section. The two $E_8$ 
singularities can be exhibited as two $\mathrm{II}^*$ fibres in
Kodaira's notation. Generically we have four $\mathrm{I}_1$ fibres
elsewhere.

Now we have the upper-left matrix $\bigl(\begin{smallmatrix}
\alpha_1&0\\0&\alpha_2\end{smallmatrix}\bigr)$ of (\ref{eq:psi})
controlling the K\"ahler form of the K3 surface and the 
lower-right submatrix $\bigl(\begin{smallmatrix}
\alpha_3&0\\0&\alpha_4\end{smallmatrix}\bigr)$ 
controlling the complex structure of the K3
surface. Thus, in the $\alpha_i\to\infty$ limit, both the K\"ahler form
and the complex structure of the K3 surface will degenerate. Note that
the choice of $\psi$ in (\ref{eq:psi}) sets the $B$-field on the K3
surface to zero. 

We will begin by focusing on the K\"ahler moduli. 
Let us begin by associating the classes generated by $H^0$ and $H^4$
to $w_1^*$ and $w_1$.
The Picard lattice
of the K3 surface is generated by the class of the section and the
class of the generic fibre. The section is a rational curve and it
thus has self-intersection $-2$. Let us call this class $C$ and
associate it to the class $w_2^*-w^2$. The generic elliptic fibre we
denote $E$ and associate it to $w_2$. It can be shown \cite{me:lK3}
that the K\"ahler form is then given by
\begin{equation}
  J = \sqrt{\frac{\alpha_1}{2\alpha_2}}\,w_2^* + 
		\sqrt{\frac{\alpha_1\alpha_2}2}\,w_2.
\end{equation}
That is, we have
\begin{equation}
\begin{split}
  \Vol(\mathrm{K3}) &= J.J = \alpha_1\\
  \Area(C) &= J.C = \sqrt{\frac{\alpha_1}{2\alpha_2}} + 
		\sqrt{\frac{\alpha_1\alpha_2}2}\\
  \Area(E) &= J.E = \sqrt{\frac{\alpha_1}{2\alpha_2}}.
\end{split}
\end{equation}
The units are those set by the string tension (i.e., the length is set
by $\sqrt{\alpha'}$).
In the limit we require we put $\alpha_2=\alpha_1$ and let $\alpha_1$
go to infinity. This produces the following behaviors
\begin{equation}
\begin{split}
\Vol(\mathrm{K3}) &\sim \alpha_1\\
\Area(C) &\sim \alpha_1\\
\Area(E) &\sim 1,  \label{eq:r1}
\end{split}
\end{equation}
That is, the area of the section of the fibration (i.e., the base)
grows while the area of the fibre remains fixed.

Now we consider the complex structure degeneration. We need to
understand how both the elliptic fibre and the section deform as
$\alpha_3$ and $\alpha_4$ are taken to infinity. We will need to use our
knowledge of the bad fibres in the elliptic fibration to achieve this
goal. At two points in the base we have type $\mathrm{II}^*$ fibres,
blown-down to produce the $E_8$ singularities in the K3 surface, and at
four other points we have $\mathrm{I}_1$ fibres. The relative
locations of these fibres move as the complex structure is varied.

One method, as explained in \cite{MV:F2}, is to attempt to analyze the
metric on the section in terms of deficit angles around the location
of the bad fibres. As shown in \cite{GSVY:cs}, if the discriminant of
the elliptic curve vanishes to order $N$ at a bad fibre, then
asymptotically one has a deficit angle of $N\pi/6$ around a bad
fibre.  One can then try to model the section in terms of curvature
localized around the locations of the bad fibres.
This method is somewhat difficult to use to obtain a quantitative
picture of the K3 degeneration in the form we require as it does not
lend itself easily to a description in terms of the moduli space we
are using.

Instead we will look at the periods of the holomorphic 2-form,
$\Omega$. The integral of $\Omega$ over a transcendental 2-cycle
(i.e., a 2-cycle that is not homologous to a holomorphic curve) will
be generically 
nonzero. One may fix a hyperk\"ahler (and thus Ricci-flat) metric on
the K3 surface while 
changing the complex structure to turn $\Omega$ into a (suitably
normalized) K\"ahler form. 
At the same time, if the transcendental cycles are suitably chosen,
they will become holomorphic curves. Thus, these periods measure the areas of
the transcendental cycles. 
The suitable transcendental 2-cycles required are the ones whose areas
are minimized within a given homology class. These are the ``supersymmetric
cycles'' as discussed in the 3-fold case in \cite{BBS:5b}. (These
object were also discussed in the K3 case in \cite{BSV:tft}.)
Knowing these areas will allow us to see how
the section and fibre degenerate in the limit we require.

The idea will be to analyze these periods in a way very analogous to
the above discussion of the areas of the algebraic cycles. This will
allow us to find a symmetry which interchanges various 2-cycles together
with the r\^oles of $\alpha_1$, $\alpha_2$, and $\alpha_3$. 

What are the transcendental 2-cycles in our elliptic fibration?
The final answer is fairly simple and has been encountered before in the
context of string theory --- see, for example, \cite{KLM:hSW}.
Technically speaking, for an elliptic fibration $\pi:S\to B$, the
cohomology of $S$ is given by the Leray spectral sequence (see, for
example, \cite{BT:}). Since the fibration has singularities we need to
be even more technical and use higher direct image sheaves (see, for
example, \cite{Hartshorne:}). Using the fact that spectral sequence
degenerates at $E_2$, we have that the second cohomology of
our K3 surface is given by the expression
\begin{equation}
  H^2(S,\Z) = H^0(B,R^2\pi_*\Z) \oplus H^1(B,R^1\pi_*\Z)
                   \oplus H^2(B,\pi_*\Z). \label{eq:ss}
\end{equation}
Roughly speaking, these objects are not as fearsome as they may first
appear. Homologically speaking (i.e., taking the 2-cycle dual to the 2-form),
the first term on the right-hand side 
of (\ref{eq:ss}) represents monodromy-invariant 2-cycles in the fibre
together with other 2-cycles coming from reducible bad fibres. The
only monodromy-invariant 2-cycle in the fibre is, of course, the
elliptic fibre itself. The last term represents the section, $C$. All
the 2-cycles in the first and final term are algebraic. Any
transcendental cycles must be in the term $H^1(B,R^1\pi_*\Z)$.

Again, roughly speaking, the term $H^1(B,R^1\pi_*\Z)$ consists of
elements dual to 2-cycles built by transporting 1-cycles in the fibre
along non-contractable 1-chains in the base. Algebraic 2-cycles in this
term correspond to elements of the \MW\ group of the
fibration (see, for example, \cite{CZ:gag}). In our case this group is
trivial and so the term $H^1(B,R^1\pi_*\Z)$ corresponds purely to our
desired transcendental 2-cycles. 

As mentioned earlier, the transcendental lattice is isomorphic to 
$\Gamma_{2,2}$. We thus need to build four 2-cycles by transporting
1-cycles in the fibre over 1-chains in the base which generate this
lattice.

The first attempt is to take a 1-cycle in the fibre and go around a
circle in the base. If this circle can be smoothly contracted then
clearly such a constructed 2-cycle will be homologically
trivial. Thus the circle must go around locations of more than one
bad fibre. In 
this case however, the 1-cycle will usually have monodromy and will not be
returned to itself. This method will only work if we can find a
collection of bad fibres such that a loop around this collection has
no monodromy on 1-cycles in the fibre. To avoid the produced 2-cycle
being homologically trivial we also demand that the monodromy
individually around each component bad fibre be nontrivial.

\iffigs
\begin{figure}[t]
  \centerline{\epsfysize=8cm\epsfbox{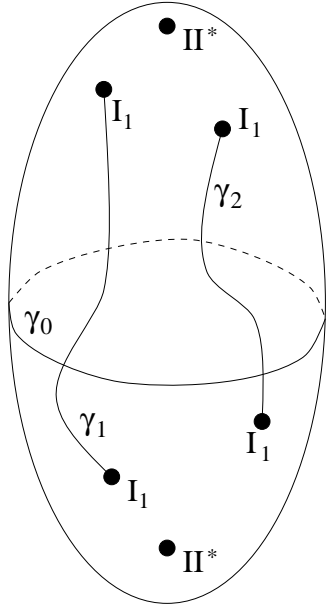}}
  \caption{One-chains producing transcendental 2-cycles.}
  \label{fig:gs}
\end{figure}
\fi

Such a path is given by $\gamma_0$ in figure \ref{fig:gs}. To see this
note that a rational elliptic surface may be constructed by taking an
elliptic fibration over $\P^1$ where the bad fibres are given by a
single type $\mathrm{II}^*$ fibre and two type $\mathrm{I}_1$
fibres. One may then take two rational elliptic surfaces, remove an
elliptic curve from each (corresponding to the anticanonical divisor)
and glue the two surfaces together along these missing curves to
produce a degenerate K3 surface. This is the stable degeneration of
the K3 surface which will be considered in more detail in the next
section. For now however we just need to note that $\gamma_0$ is the
circle that divides the bad fibres into the two sets that can each
produce a rational elliptic surface. The fact that these sets of bad
fibres can produce a good elliptic surface by themselves must mean
that the monodromy around them is trivial.

Either of the 1-cycles that generate $H_1$ of the fibre may be
transported around $\gamma_0$ and each give a 2-torus. Thus we have
produced two transcendental 2-tori. Let us call these tori
$\Scr{E}_1$ and $\Scr{E}_2$.

The other two 2-cycles must be built differently. Consider two
$\mathrm{I}_1$ locations in the base brought together so that they
form coalesce and produce an $\mathrm{I}_2$ fibre. This fibre must be
blown-up to produce a smooth elliptic surface. This blow-up produces
a new rational curve. Since the total cohomology of the surface cannot
have changed during this process, there must have been a
transcendental 2-sphere which shrank down to zero size as the two 
$\mathrm{I}_1$ locations approached each other. It is easy to see how
to make such a 2-sphere. An $\mathrm{I}_1$ fibre is exactly the stable
degeneration of a 2-torus discussed in section \ref{s:T2} in which one
1-cycle of the fibre shrinks down to zero size. In our case we have
two $\mathrm{I}_1$ fibres in which the same cycle shrinks (otherwise
they would not coalesce to form an $\mathrm{I}_2$ fibre). The 2-sphere
can therefore be built as parallel lines of ``latitude'' on the
2-sphere given by this 1-cycle in the fibre which shrinks down to
zero at the north and south poles. We show this in figure \ref{fig:tc}.

\iffigs
\begin{figure}[t]
  \centerline{\epsfxsize=10cm\epsfbox{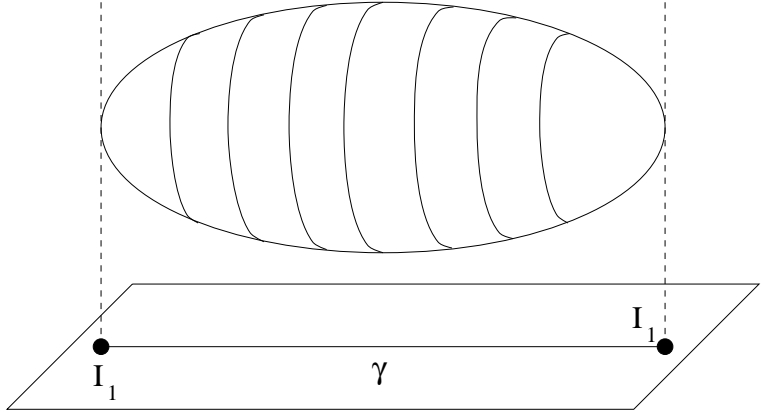}}
  \caption{Building a transcendental 2-sphere.}
  \label{fig:tc}
\end{figure}
\fi

We may thus associate a transcendental 2-sphere to the path between
two $\mathrm{I}_1$ fibres that would produce a type $\mathrm{I}_2$
fibre if they were to coalesce. This need not be the case for any pair
of type $\mathrm{I}_1$ fibres however. Consider the upper pair of
$\mathrm{I}_1$ fibres in figure \ref{fig:gs}. We know that these two
fibres, together with the upper $\mathrm{II}^*$ fibre can produce a
rational elliptic curve. The elliptic surface produced by a
$\mathrm{II}^*$ fibre and a type $\mathrm{I}_2$ fibre would have Picard
number 11 (one from the section, one from the generic fibre, eight
more from the $\mathrm{II}^*$ fibre and one more from the
$\mathrm{I}_2$ fibre). As the rational elliptic surface has Picard
number 10 this is impossible. 
Each of these type $\mathrm{I}_1$ fibres
contracts a different circle in the fibre. The same is true for the
lower pair of $\mathrm{I}_1$ fibres.

It is not hard to convince oneself that the only way to produce the
desired result is to generate two 2-spheres by using the paths shown
as $\gamma_1$ and $\gamma_2$ in figure \ref{fig:gs}. Let us call these
2-cycles $\Scr{C}_1$ and $\Scr{C}_2$ respectively. One may show that the
intersection numbers are given by
\begin{equation}
\begin{split}
  \Scr{E}_i\cdot\Scr{E}_j &= 0\\
  \Scr{E}_i\cdot\Scr{C}_j &= \delta_{ij}\\
  \Scr{C}_i\cdot\Scr{C}_j &= -2\delta_{ij},
\end{split}
\end{equation}
thus generating $\Gamma_{2,2}$ as desired.

As we saw earlier, the 2-torus, $E$, together with the 2-sphere, $C$
generate the Picard lattice, $\Gamma_{1,1}$. In addition we now see
that the 2-torus,
$\Scr{E}_1$, together with the 2-sphere, $\Scr{C}_1$, generate
another $\Gamma_{1,1}$ while $\Scr{E}_2$ and $\Scr{C}_2$ generate a
third $\Gamma_{1,1}$. 

We have now given an explicit construction of the lattice
$\Gamma_{4,4}$ in (\ref{eq:dec1}) as
\begin{equation}
  \Gamma_{4,4} =
\Gamma_{1,1}\oplus\Gamma_{1,1}\oplus\Gamma_{1,1}\oplus\Gamma_{1,1}.
	\label{eq:dec4}
\end{equation}
The first $\Gamma_{1,1}$ term can be associated to the quantum, $H_0$
and $H_4$, part of the homology. The second term is the Picard lattice
and the third and fourth terms are generated by the transcendental
cycles $\{\Scr{C}_1,\Scr{E}_1\}$ and $\{\Scr{C}_2,\Scr{E}_2\}$
respectively. The parameters $\alpha_1,\alpha_2,\alpha_3$, and
$\alpha_4$ can also be associated to each of the four terms in 
(\ref{eq:dec4}). Thus by setting these parameters equal, there will be
a symmetry of the K3 surface that interchanges these sets.
That is, fixing a hyperk\"ahler 
metric on the K3 surface, but allowing the complex structure to vary,
we may take any of these three $\Gamma_{1,1}$ 
lattices to be the Picard lattice. 

Since we set $\alpha_2=\alpha_3=\alpha_4$ any statement about the
areas of the 2-cycles $C$ and $E$ must also be true about the 2-cycles
$\Scr{E}_i$ and $\Scr{C}_i$. That is from (\ref{eq:r1}), in our limit,
\begin{equation}
\begin{split}
\Area(\Scr{C}_i) &\sim \alpha_1\\
\Area(\Scr{E}_i) &\sim 1,
\end{split}
\end{equation}
for $i=1$ or $2$. 

We now want to know how to relate the areas of the supersymmetric
cycles to the lengths of the 1-cycles we have drawn in figure
\ref{fig:gs}. Since $\pi_1(\mathrm{K3})=0$, we have that 
$H_2(\mathrm{K3})=\pi_2(\mathrm{K3})$. To build a supersymmetric
cycle we can therefore continuously shrink the transcendental cycles
we have built above. We will take it as being intuitively clear that
the resulting minimal area will be proportional to the product of the shortest
possible 1-cycle in the fibre and shortest possible 1-chain in the
base in the right class.\footnote{We are not being particularly
rigorous here. One might be able to construct the supersymmetric
cycles explicitly along the lines of \cite{Bry:lag} and then tie it to
our construction but we will not attempt this here. I thank
D.~Morrison for pointing this out to me.}

That is, up to factors which we are ignoring, the area of a 2-cycle
generated by transporting a 1-cycle in the fibre over a one-chain in
the base is given by a product of the length of the 1-cycle in
the fibre times the length of the one-chain in the base. The only way
to achieve all the areas to be the desired size is to make both cycles
in the generic fibre of length order one and the chains in the base to satisfy
\begin{equation}
\begin{split}
\mathrm{Length}(\gamma_0) &\sim 1\\
\mathrm{Length}(\gamma_1) &\sim \alpha_1\\
\mathrm{Length}(\gamma_2) &\sim \alpha_1,
\end{split}
\end{equation}
In the above we assume each chain above is some kind of ``geodesic'' of
shortest possible length in its class.
The only way to achieve this is to stretch figure \ref{fig:gs} into a
long vertical sausage of length $\alpha_1$.

The picture as $\alpha_1\to\infty$ is therefore of a K3 surface which
is stretching out in one direction but remaining fixed in the other
three directions. When written as an elliptic fibration, the generic
fibre remains fixed in size but the section (or base) elongates in one
direction. 

Therefore if we take an $E_8\times E_8$ heterotic string on a
four-torus and allow the four-torus to dilate equally in all four
directions, the dual type IIA string is compactified on a space which
is only large in one dimension. It is interesting to compare this with
the calculation in \cite{W:dyn} which showed that if the type IIA
string's K3 surface is dilated in all four directions equally, then
the heterotic four-torus expands in only one direction, leaving a fixed
three-torus. This latter correspondence was used to show that M-theory
on a K3 surface is dual to the heterotic string on a three-torus. Our
case is somewhat different as we wish to consider the final dual pair
in ten dimensions rather than seven.

Let us denote the ten-dimensional heterotic dilaton by 
$\lambda_{\mathrm{H},10}$ and the ten-dimensional type II dilaton by
$\lambda_{\mathrm{II},10}$. Similarly, the six-dimensional dilatons
are given by $\lambda_{\mathrm{H},6}$ and $\lambda_{\mathrm{II},6}$,
respectively. We also have metrics $g_{\mathrm{H}}$ and
$g_{\mathrm{II}}$ by which lengths are measured.
We need to be a little careful in the following argument
to keep track of the string scale correctly. We use $L$ to denote the
natural type IIA scale. That is, our K3 surface above will be of
length $L$ in three directions. Now we have
\begin{equation}
  \lambda_{\mathrm{H},10}^2 = \Vol_{\mathrm{H}}(T^4)\lambda_{\mathrm{H},6}^2
    \sim \alpha_1^2\lambda_{\mathrm{H},6}^2,
\end{equation}
and
\begin{equation}
  \lambda_{\mathrm{II},10}^2 = \Vol_{\mathrm{II}}(\mathrm{K3})
     \lambda_{\mathrm{II},6}^2
    \sim \alpha_1L^3\lambda_{\mathrm{II},6}^2.
\end{equation}
Heterotic -- type II duality however dictates that
\begin{equation}
  \lambda_{\mathrm{H},6} = \lambda_{\mathrm{II},6}^{-1},
\end{equation}
and so
\begin{equation}
  \lambda_{\mathrm{H},10}^2 \sim
    \frac{\alpha_1^3L^3}{\lambda_{\mathrm{II},10}^2}.
\end{equation}
This shows that as $\alpha_1\to\infty$, the only way that we may
maintain a finite coupling for the ten-dimensional heterotic string is
to let the type IIA string become infinitely coupled --- that is, we
require M-theory. Now, as shown in \cite{W:dyn}, the metric in
M-theory is scaled with respect to the type IIA string according to
\begin{equation}
  g_{\mathrm{M}} = g_{\mathrm{II}}\lambda_{\mathrm{II},10}^{-\frac23}.
\end{equation}
Therefore, as $\lambda_{\mathrm{II},10}\to\infty$ M-theory lengths are
infinitely shorter than type II lengths. This means that whereas the
type IIA string's K3 surface was expanding in one direction, the
M-theory K3 surface will actually shrink in the three dimensions. 
The resulting shape is therefore a line segment which we denote
$\mathcal{I}$. This 
is the ``squashed K3 surface'' of \cite{AM:Ud}. To be precise, put
\begin{equation}
  \ell = \alpha_1\lambda_{\mathrm{II},10}^{-\frac13},
\end{equation}
as the length of the $\gamma_1$ and $\gamma_2$ lines in M-theory
units, i.e., the length of the squashed K3 surface as measured by
M-theory. We also put $L=\lambda_{\mathrm{II},10}^{\frac13}$. Thus we
obtain
\begin{equation}
  \lambda_{\mathrm{H},10}^2 \sim \ell^3.
\end{equation}
This is in complete agreement with the Ho\v{r}ava--Witten picture
\cite{HW:E8M} which uses quite different logic and describes $\mathcal{I}$
as $S^1/\Z_2$.

Note that the remnants of the $E_8$ structure in the original K3
surface have been squeezed down to points. Thus the geometry of this
structure has been lost completely. M-theory itself must therefore
just assert that an $E_8$ Yang--Mills theory lives at each of the ends
of the line segment without giving it a geometrical manifestation. In
\cite{HW:E8M} the existence of the $E_8$'s was shown by anomaly arguments.


\section{F-Theory}   \label{s:F}

The degeneration of the K3 surface in the previous section was clearly
analogous to the degeneration in section \ref{s:T2} of the 2-torus 
which collapsed onto a circle. That is, every 1-cycle in a given class
is shrunk down to zero size resulting in a dimension change from two
to one.

In this section we will pursue the algebraic geometer's ``stable
degeneration'' analogous to just shrinking down one cycle of the
two-torus. This was 
proposed in \cite{FMW:F} and then explained in length in \cite{AM:po}.
We will not repeat that description here but just emphasize certain
points. We thus refer the reader to section 3 of \cite{AM:po} for
details regarding some of the discussion below.

The first stage of the decompactification in the previous section
consisted of stretching the K\"ahler form so that the section of the
K3 surface had an area much greater than the elliptic fibre. We do the
same for F-theory. In fact this may be taken as the defining step that
takes the type IIA string to F-theory. Again, as in M-theory, one may
speculate that F-theory may exist in its own right, rather than just
being a limit of string theory. Again we will not pursue such a line of
reasoning here (and it seems less likely than for M-theory).

The F-theory statement is therefore an eight-dimensional one in which
one asserts that F-theory on a K3 surface is dual to the heterotic
string on $T^2$ \cite{Vafa:F}. In contrast to the previous section,
this $T^2$ will not be 
decompactified. What we can do however is to ask what happens when
this $T^2$ becomes very large. This produces the stable degeneration
for the F-theory K3 surface.

This may appear to be quite a subtle point as to whether we consider
the $T^2$ as being decompactified or whether we consider it to be
taken to be very large. What is clear however is that it does lead to
quite different pictures. For example, the complex structure of this
$T^2$ remains as a modulus if we merely take the $T^2$ to be large. If
we decompactify, this modulus is lost as $\R^2$ has no complex
deformation.

The complex degeneration of the K3 surface we require therefore is one
that preserves more geometric structure than the one considered in
section \ref{s:M}. By analogy with the 2-torus of section \ref{s:T2}
it is clear what we may do. We saw in section \ref{s:M} that the
periods associated to line $\gamma_0$ of figure \ref{fig:gs} were very
small relative to the periods associated to those of $\gamma_1$ and
$\gamma_2$. The M-theory picture of this was to shrink all the cycles
in the class of $\gamma_0$ down to zero size. This squeezed the
2-sphere in figure \ref{fig:gs} down to a line segment. Such a
procedure was forced upon us if we wished to respect the Ricci-flat
metric --- i.e., the hyperk\"ahler structure. 

\iffigs
\begin{figure}[t]
  \centerline{\epsfxsize=11cm\epsfbox{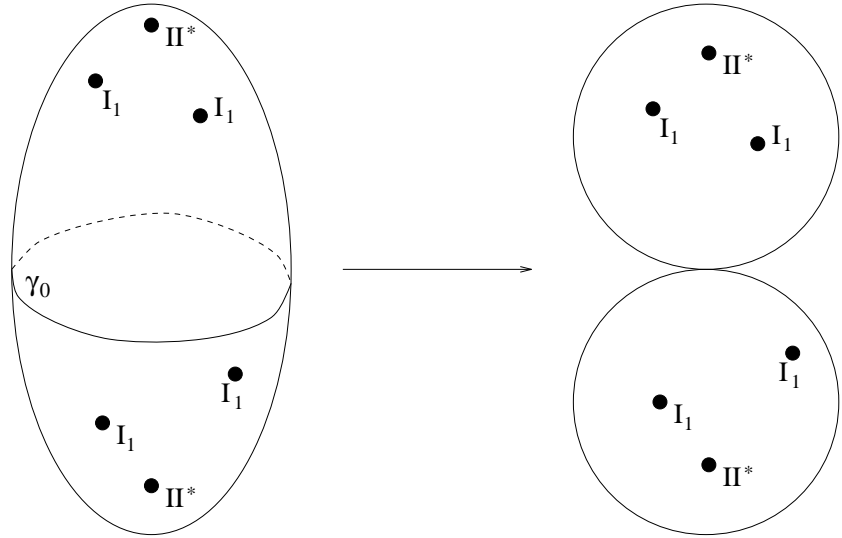}}
  \caption{The stable degeneration of an elliptic K3.}
  \label{fig:sd}
\end{figure}
\fi

If we forget about metrics, we are free to shrink the single cycle
$\gamma_0$ down to zero size producing figure \ref{fig:sd}. This changes the
base of the elliptic fibration from a single $\P^1$ to two $\P^1$'s
intersecting at a point. Therefore, the K3 surface itself degenerates
into two rational elliptic surfaces intersecting along an elliptic
curve. This latter elliptic curve is the fibre over the point of
intersection of the two base $\P^1$'s. This is the stable degeneration
of \cite{FMW:F,AM:po} and is clearly the analogue of the stable
degeneration of the 2-torus in section \ref{s:T2}.

It is striking to see just how different the two pictures of the
degeneration are. The M-theory degeneration consisted of a line
segment with little structure. The F-theory degeneration consists of
two rational elliptic surfaces. 
Again, we should emphasize that the F-theory picture is not a
complete decompactification. The 
complex structure of the $T^2$ remains as a modulus. In fact, as
explained in \cite{FMW:F,AM:po} the modulus of the $T^2$ on which the
$E_8\times E_8$ heterotic string is compactified is exactly the
modulus of the elliptic curve which is the intersection of the two
rational elliptic surfaces on the F-theory side. This elliptic curve,
which we will call $E_*$, is the fibre over the point of intersection
over the two $\P^1$'s in figure \ref{fig:sd}.
Note also that while the $E_8$ structures were
squeezed out of geometry in M-theory, they are still very much
present in the F-theory degeneration in the rational elliptic surfaces.

In order to compactify the $E_8\times E_8$ heterotic string on $T^2$,
we need to choose an instanton background for each of the two $E_8$'s
on the 2-torus. The moduli space of such objects is (up to some
subtleties of $S$-equivalence) the same as the moduli space of
semi-stable holomorphic bundles with structure group given by the
complex group $E_8$. It was explained in \cite{FMW:F} (see also the
appendix of \cite{ BJPS:F}) that such
bundles could be built from $E_8$ del Pezzo and that the moduli space of
such del Pezzo surfaces yielded the desired moduli space.

In our case we have rational elliptic surfaces representing the
$E_8$-bundles rather than del Pezzo surfaces. This is not much of a
problem however. A rational elliptic surface can be written as an
$E_8$ del Pezzo surface blown up at one point. Indeed, the model of
the stable degeneration of the K3 surface in question can be flopped
so as to replace the rational elliptic surfaces with $E_8$ del Pezzo
surfaces.\footnote{Actually the construction of \cite{FMW:F} can be
modified so that it uses the rational elliptic surface directly. In
this case the spectral data is built from intersections of elements of
the Mordell--Weil group of the rational elliptic curve with
$E_*$. This construction produces a rather direct picture of the vector
bundle data in the stable degeneration and should be studied further.}


\section{M-theory from F-theory}   \label{s:MF}

We have two pictures of the $E_8\times E_8$ heterotic string on an
elliptic curve, $E_*$. The F-theory picture, in the large $E_*$ limit,
is that of the stable 
degeneration in figure \ref{fig:sd} where $E_*$ is the fibre (at least
as far as complex structure is concerned) over the
point of intersection of the two base $\P^1$'s. The M-theory picture
is a compactification on $\mathcal{I}\times E_*$. It is clear how to
produce the M-theory picture from the limiting F-theory picture --- simply
squeeze the base $\P^1$'s in figure \ref{fig:sd} to produce
$\mathcal{I}$. That is, modify figure
\ref{fig:sd} so that all of the cycles in the 
class $\gamma_0$ shrink to zero size rather that just one. 

This should allow the
M-theory picture of any F-theory compactification to be constructed.
We will illustrate this procedure here by considering point-like
$E_8$-instantons from the $E_8\times E_8$ heterotic string
compactified on a K3 surface. 

In order to compactify the $E_8\times E_8$ heterotic string on a K3
surface one requires two $E_8$-instanton backgrounds on the K3 surface
in order to compactify the gauge degrees of freedom of the heterotic
string. In the simplest case, 24 instantons are required to cancel
anomalies. These instantons have a moduli space and lead to moduli in
the string compactification. These moduli live in hypermultiplets of
the uncompactified six-dimensional theory.
The branch of the moduli space consisting
of such variations is conventionally called the ``Higgs'' branch. It
was shown in \cite{GH:E8i,SW:6d} that if one of these instantons becomes
point-like in the K3 surface then new massless moduli may appear
leading to a phase transition. These moduli are in tensor multiplets
of the six-dimensional field theory and form the ``Coulomb branch''.

The approach used in \cite{GH:E8i,SW:6d} was based on M-theory compactified
on $\mathcal{I}\times$K3. This is the picture we wish to
``derive''. An alternative picture was given in \cite{MV:F,MV:F2}
based on F-theory. This is the picture we will take as the starting
point and which we now review briefly. We refer to \cite{me:lK3} for an
pedagogical description.

Consider F-theory on a \CY\ threefold, $X$, which is an elliptic
fibration over a Hirzebruch surface $\HS n$. Let $f$ denote a generic
$\P^1$-fibre of the Hirzebruch surface and let $C_0$ denote the
section with self-intersection $-n$. Restricting the elliptic
fibration to any $f$ produces a K3 surface. Thus $X$ is also a
K3-fibration over $\P^1$. 

Unbroken nonabelian gauge groups can be generated by curves of certain
fibres within $\HS n$ and perturbative gauge symmetries arise from
sections of $\HS n$ (or multi-sections for higher-level symmetries). We
may model all 24 instantons as being small by forcing an $E_8\times
E_8$ unbroken gauge symmetry which is produced by two sections of
$\mathrm{II}^*$ fibres in $\HS n$. When this is done, the geometry
allows one to blow-up the base $\HS n$ while keeping the elliptic
fibration \CY. This corresponds to the phase transition produced by
the small $E_8$-instantons.

\iffigs
\begin{figure}[t]
  \centerline{\epsfxsize=11cm\epsfbox{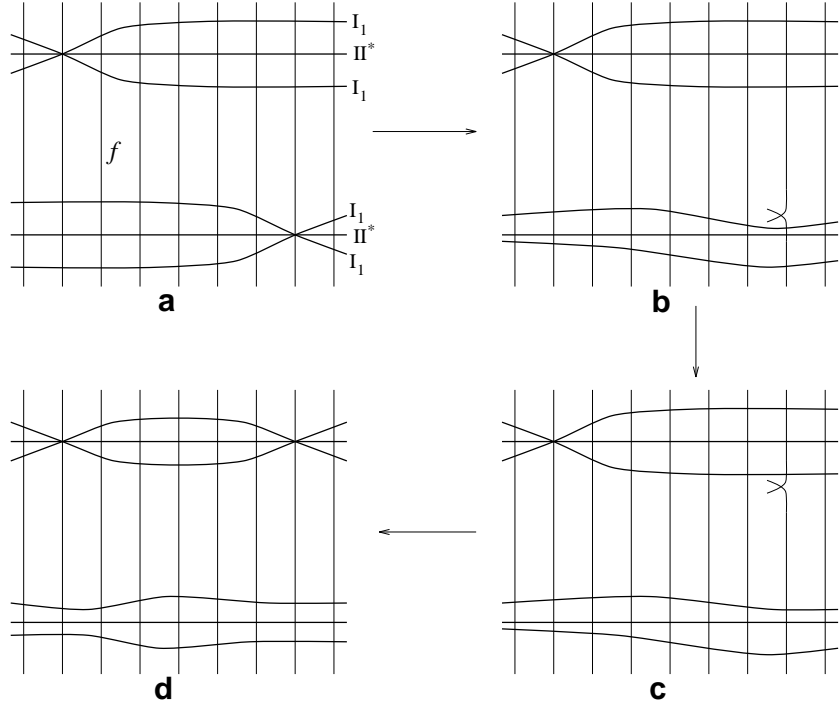}}
  \caption{The F-theory picture of moving point-like $E_8$-instantons.}
  \label{fig:tran}
\end{figure}
\fi

This precise geometry of such a transition is shown in figure
\ref{fig:tran}. In this diagram complex dimensions are drawn as real,
i.e., complex curves are drawn as real curves.
The horizontal lines and curves denote complex curves of singular
fibres of the Kodaira type as indicated. The vertical
lines represent $\P^1$-fibres, $f$, of the Hirzebruch surface (over
which the elliptic fibre is generically smooth). Figure
\ref{fig:tran}a shows two point-like instantons --- one in one $E_8$
and one in the other $E_8$, each represented in F-theory by collisions
of curves of $\mathrm{II}^*$ fibres with curves of $\mathrm{I}_1$ as
shown. Such a collision can be blown-up yielding the transition into
the Coulomb branch. This is done for the bottom $E_8$-instanton in
figure \ref{fig:tran}b. This produces a new curve in the base. The
effect of this is to replace the $f$-line that had passed through the
location of the instanton by two $\P^1$'s intersecting
transversely. If we just blow-up a little bit, the new $\P^1$ that passes
through the former location of the instanton will be small while the
other $\P^1$ will be only slightly smaller than the generic $f$-line. 

One may increase the size of the blow-up by changing the relative
sizes of the two $\P^1$'s. Note that by homology arguments the total
area of the these two 
$\P^1$'s is fixed and equal to the area of the generic $f$-line. The
larger blow-up is shown in figure \ref{fig:tran}c. We cannot increase
the size of the blow-up indefinitely --- eventually the original
$\P^1$ will shrink down to zero size. This is shown in figure
\ref{fig:tran}d and corresponds to the point-like instanton having
moved from one $E_8$ to the other.

Each $f$-line represents the base of an elliptic K3 surface on which
the F-theory is compactified. We know however that we can turn such a
K3 surface into the M-theory picture by squashing this $\P^1$ down
to a line segment. To be exact, we should first go to the stable
degeneration, which picks out a special $T^2$ fibre in the elliptic
fibration, which we called $E_*$. Then we squash the $\P^1$ down
completely to reproduce M-theory on $\mathcal{I}\times E_*$ for each $f$-line.

When we go to the stable degeneration, $E_*$ will vary as a move
around the base $\P^1$ of the Hirzebruch surface.
The geometry of the stable degeneration was discussed in
\cite{AM:po}. The result is that $E_*$ acts as the elliptic fibre of a
K3 surface as we move over the base $\P^1$. This means that globally
the picture becomes M-theory compactified on $\mathcal{I}\times$K3.

This is story for the generic $f$-lines at least. We should also worry
about the special fibres in figures \ref{fig:tran}b and
\ref{fig:tran}c which have become two $\P^1$'s. What happens to these
as we go to the M-theory limit?

Actually squashing these two $\P^1$'s is exactly the same as squashing
the stable degeneration picture of the K3 surface. The equivalence
between the geometry of stable degenerations and point-like instantons
was discussed in \cite{AM:po}. 
We are thus considering the same degeneration as at the start of this section.
That is, each $\P^1$ squashes down to a line and then these two lines
are joined end-to-end to form a longer line. 
This longer line will be the same length as the generic
$\mathcal{I}$'s produced by generic $f$-lines.
It is important to note
however that the point where these two lines join is significant. As
we vary the size of the blow-up, this point will move up and down the
resulting line $\mathcal{I}$.

\iffigs
\begin{figure}[t]
  \centerline{\epsfxsize=10cm\epsfbox{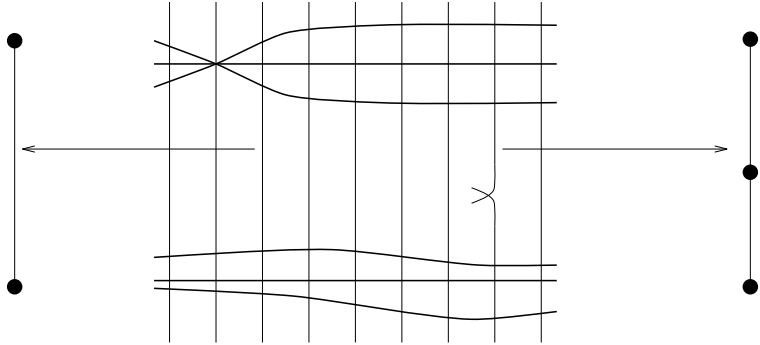}}
  \caption{Squashing $f$-lines into $\mathcal{I}$-lines.}
  \label{fig:MF}
\end{figure}
\fi

This process of moving from F-theory to M-theory is shown in figure
\ref{fig:MF}. The vertical {\em complex\/} $f$-lines in the picture in
F-theory are squeezed into {\em real\/} line segments of the form
$\mathcal{I}$. The reducible fibres corresponding to blow-ups induced
by point-like instantons result in a point being marked within the
relevant line segments.

The result then is that figure \ref{fig:tran}, in the M-theory limit,
will be an $(\mathcal{I}\times E_*)$-bundle over $\P^1$, except at the
image of the locations of the point-like instantons in
the base $\P^1$ where the $\mathcal{I}$ will have a special
marked point. It is easy to see how to match this up to the
picture of \cite{GH:E8i,SW:6d}. This special ``marked point''
on $\mathcal{I}$ is the location of a ``5-brane''. The transition
shown in figure \ref{fig:tran} turns into a picture of a 5-brane
moving from one end of $\mathcal{I}$ to the other end.

Note that in our description we have only specified the location of
the 5-brane over the base $\P^1$ and not as a point in the K3
surface. To fix a point in the K3 surface we need to further localize
the 5-brane to a point in the $E_*$-fibre. This degree of freedom is
a hypermultiplet degree of freedom and is of the type usually
``missed'' by F-theory. See \cite{FMW:F} for further discussion of
these extra moduli in the hypermultiplets.

We see then that the F-theory picture of point-like instantons can be
mapped in a very direct fashion to the M-theory picture. 


\section*{Acknowledgements}

It is a pleasure to thank O.~Aharony, M.~Gross, S.~Kachru, D.~Morrison, and
E.~Silverstein for useful conversations.  The author is supported by
DOE grant DE-FG02-96ER40959.


\begin{thebibliography}{10}

\bibitem{Sen:uni}
A.~Sen,
\newblock {\em Unification of String Dualities},
\newblock hep-th/9609176.

\bibitem{BFSS:M}
T.~Banks, W.~Fischler, S.~H. Shenker, and L.~Susskind,
\newblock {\em M Theory as a Matrix Model: A Conjecture},
\newblock Phys. Rev. {\bf D55} (1997) 5112--5128, hep-th/9610043.

\bibitem{MV:F2}
D.~R. Morrison and C.~Vafa,
\newblock {\em Compactifications of F-Theory on Calabi--Yau Threefolds --- II},
\newblock Nucl. Phys. {\bf B476} (1996) 437--469, hep-th/9603161.

\bibitem{FMW:F}
R.~Friedman, J.~Morgan, and E.~Witten,
\newblock {\em Vector Bundles and F Theory},
\newblock hep-th/9701162,
\newblock (revised version).

\bibitem{CJ:MF}
C.~V. Johnson,
\newblock {\em From M-theory to F-theory, with Branes},
\newblock hep-th/9706155.

\bibitem{me:lK3}
P.~S. Aspinwall,
\newblock {\em K3 Surfaces and String Duality},
\newblock hep-th/9611137,
\newblock to appear in the proceedings of TASI 96.

\bibitem{HW:E8M}
P.~Ho{\v r}ava and E.~Witten,
\newblock {\em Heterotic and Type I String Dynamics from Eleven Dimensions},
\newblock Nucl. Phys. {\bf B460} (1996) 506--524, hep-th/9510209.

\bibitem{GH:E8i}
O.~J. Ganor and A.~Hanany,
\newblock {\em Small $E_8$ Instantons and Tensionless Noncritical Strings},
\newblock Nucl. Phys. {\bf B474} (1996) 122--140, hep-th/9602120.

\bibitem{SW:6d}
N.~Seiberg and E.~Witten,
\newblock {\em Comments on String Dynamics in Six Dimensions},
\newblock Nucl. Phys. {\bf B471} (1996) 121--134, hep-th/9603003.

\bibitem{FM:BGOD}
R.~Friedman and D.~R. Morrison, editors,
\newblock {\em The Birational Geometry of Degenerations}, volume~29 of Progress
  in Math.,
\newblock Birkh{\"a}user, Boston, Basel, Stuttgart, 1983.

\bibitem{AM:Ud}
P.~S. Aspinwall and D.~R. Morrison,
\newblock {\em U-Duality and Integral Structures},
\newblock Phys. Lett. {\bf 355B} (1995) 141--149, hep-th/9505025.

\bibitem{BR:K3M}
M.~Berkooz and M.~Rozali,
\newblock {\em String Dualities from Matrix Theory},
\newblock hep-th/9705175.

\bibitem{Gov:K3M}
S.~Govindarajan,
\newblock {\em A Note on M(atrix) Theory in Seven Dimensions with Eight
  Supercharges},
\newblock hep-th/9705113.

\bibitem{AM:K3p}
P.~S. Aspinwall and D.~R. Morrison,
\newblock {\em String Theory on K3 Surfaces},
\newblock in B.~Greene and S.-T. Yau, editors, ``Mirror Symmetry II'', pages
  703--716, International Press, 1996,
\newblock hep-th/9404151.

\bibitem{GSVY:cs}
B.~R. Greene, A.~Shapere, C.~Vafa, and S.-T. Yau,
\newblock {\em Stringy Cosmic Strings and Noncompact Calabi--Yau Manifolds},
\newblock Nucl. Phys. {\bf B337} (1990) 1--36.

\bibitem{BBS:5b}
K.~Becker, M.~Becker, and A.~Strominger,
\newblock {\em Five-branes, Membranes and Nonperturbative String Theory},
\newblock Nucl. Phys. {\bf B456} (1995) 130--152, hep-th/9507158.

\bibitem{BSV:tft}
M.~Bershadsky, V.~Sadov, and C.~Vafa,
\newblock {\em D-branes and topological field theories},
\newblock Nucl. Phys. {\bf B463} (1996) 420--434, hep-th/9511222.

\bibitem{KLM:hSW}
A.~Klemm et~al.,
\newblock {\em Self-Dual Strings and N=2 Supersymmetric Field Theory},
\newblock Nucl. Phys. {\bf B477} (1996) 746--766, hep-th/9604034.

\bibitem{BT:}
R.~Bott and L.~W. Tu,
\newblock {\em Differential Forms in Algebraic Topology},
\newblock Springer-Verlag, New York, 1982.

\bibitem{Hartshorne:}
R.~Hartshorne,
\newblock {\em Algebraic Geometry}, volume~52 of Graduate Texts in Mathematics,
\newblock Springer-Verlag, 1977.

\bibitem{CZ:gag}
D.~A. Cox and S.~Zucker,
\newblock {\em Intersection Numbers of Sections of Elliptic Surfaces},
\newblock Invent. Math. {\bf 53} (1979) 1--44.

\bibitem{Bry:lag}
R.~L. Bryant,
\newblock {\em Minimal Lagrangian Submanifolds of K{\"a}hler--Einstein
  Manifolds},
\newblock in ``Differential Geometry and Differential Equations, Shanghai
  1985'', volume 1255 of Lecture Notes in Math., pages 1--12, Springer-Verlag,
  1985.

\bibitem{W:dyn}
E.~Witten,
\newblock {\em String Theory Dynamics in Various Dimensions},
\newblock Nucl. Phys. {\bf B443} (1995) 85--126, hep-th/9503124.

\bibitem{AM:po}
P.~S. Aspinwall and D.~R. Morrison,
\newblock {\em Point-like Instantons on K3 Orbifolds},
\newblock hep-th/9705104.

\bibitem{Vafa:F}
C.~Vafa,
\newblock {\em Evidence for F-Theory},
\newblock Nucl. Phys. {\bf B469} (1996) 403--418, hep-th/9602022.

\bibitem{BJPS:F}
M.~Bershadsky, A.~Johansen, T.~Pantev, and V.~Sadov,
\newblock {\em On Four-Dimensional Compactifications of F-Theory},
\newblock hep-th/9701165.

\bibitem{MV:F}
D.~R. Morrison and C.~Vafa,
\newblock {\em Compactifications of F-Theory on Calabi--Yau Threefolds --- I},
\newblock Nucl. Phys. {\bf B473} (1996) 74--92, hep-th/9602114.

\end{thebibliography}

\end{document}